\documentclass{article}

\usepackage{PRIMEarxiv}

\usepackage[utf8]{inputenc} 
\usepackage[T1]{fontenc}    
\usepackage{hyperref}       
\usepackage{url}            
\usepackage{booktabs}       
\usepackage{amsfonts}       
\usepackage{nicefrac}       
\usepackage{microtype}      
\usepackage{lipsum}
\usepackage{fancyhdr}       
\usepackage{graphicx}       
\graphicspath{{media/}}     

\usepackage[utf8]{inputenc} 
\usepackage[T1]{fontenc}  
\usepackage{amssymb}
\usepackage{booktabs}       
\usepackage{amsfonts}       
\usepackage{nicefrac}       
\usepackage{microtype}
\usepackage{bm} 
\usepackage{lipsum}
\usepackage{fancyhdr}       
\usepackage{graphicx}       
\graphicspath{{media/}}     

\usepackage[english]{babel}
\usepackage[utf8]{inputenc}
\usepackage{amsmath}
\usepackage{tikz,pgfplots}
\usepackage{tikz-network}
\usetikzlibrary{shapes,decorations,arrows,calc,arrows.meta,fit,positioning}
	
\usepackage{color, colortbl}
	
\definecolor{Gray}{gray}{0.9}
\title{Causal Discovery and Counterfactual Recommendations for Personalized Student Learning
\thanks{\textit{\underline{Citation}}: 
\textbf{Authors. Title. Pages.... DOI:000000/11111.}} 
}

\author{
  Bevan I. Smith \\
  School of Mechanical, Industrial and Aeronautical Engineering\\
  University of the Witwatersrand \\
    Johannesburg, South Africa\\
  \texttt{\ bevan.smith@wits.ac.za}} 

\begin{document}
\maketitle

\begin{abstract}

The paper focuses on identifying the causes of student performance to provide personalized recommendations for improving pass rates. We introduce the need to move beyond predictive models and instead identify causal relationships. We propose using causal discovery techniques to achieve this. The study's main contributions include using causal discovery to identify causal predictors of student performance and applying counterfactual analysis to provide personalized recommendations.

The paper describes the application of causal discovery methods, specifically the PC algorithm, to real-life student performance data. It addresses challenges such as sample size limitations and emphasizes the role of domain knowledge in causal discovery. The results reveal the identified causal relationships, such as the influence of earlier test grades and mathematical ability on final student performance.

Limitations of this study include the reliance on domain expertise for accurate causal discovery, and the necessity of larger sample sizes for reliable results. The potential for incorrect causal structure estimations is acknowledged. A major challenge remains, which is the real-time implementation and validation of counterfactual recommendations. 

In conclusion, the paper demonstrates the value of causal discovery for understanding student performance and providing personalized recommendations. It highlights the challenges, benefits, and limitations of using causal inference in an educational context, setting the stage for future studies to further explore and refine these methods.

\end{abstract}


\section{Introduction}
What causes a student to be at-risk of failing? By knowing the causes of student performance, we are able to provide personalized recommendations to improve pass rates and throughput.  In terms of student performance, much research has been carried out using machine learning to predict student performance \cite{Kotsiantis2004,Kotsiantis2010}, and although predicting student performance is indeed valuable, we do not only want to know who is predicted to be at-risk, but what is causing the poor performance.   

One potential way is to perform interpretable (or explainable) machine learning \cite{Molnar2019}. This is to identify the most significant predictors of the outcome \cite{BevanSmith2022a}: to explain why a student is at-risk \cite{Molnar2019}.  The limitation of explainable machine learning is that we cannot assume the predictors to be \textit{causal} to the outcome, only correlated.  Explainable machine learning can describe the student well, but we cannot assume those explanations to be causal. 

To answer causal questions and determine if predictors (or variables) are causal, we need causal inference and causal discovery.  To provide recommendations for an at-risk student on how to change some aspect of his/her studies in order to pass, it is vital that we identify causal features in the data and not only correlated features.  Only causal features can affect the outcome of the student's performance.  

This study contributes two main ideas: using \textbf{causal discovery} to identify causal predictors of student performance and applying \textbf{counterfactual analysis} to show how to change causal features to achieve a desirable output for a student.   We show how to take a real-life dataset and use causal discovery to estimate the true causal structure in the data which then allows us to generate counterfactual scenarios to provide personal recommendations for students.

\section{Causal Inference and Discovery}
Causal inference is the field of finding causes of things.  Judea Pearl defines causality as follows:  A causes B if B \textit{listens} to A \cite{Pearl2017}.  This means that if we change A, then we observe a change in B as well. A does not need to be a direct cause of B, but can also be a cause of events that cause B.

\subsection{Causal Graphs}
\begin{figure}[h!]
\centering
\renewcommand{\figurename}{\textbf{Fig.}}
\begin{tikzpicture}
\Vertex[x=1,size = 0.7,label = $X$, fontscale = 2]{X}
\Vertex[size =0.7, x=5,label = $y$, fontscale =2.5]{Y}
\Vertex[size=0.7,x=3,y=2,fontscale = 2,label = $Z$]{Z}
\Edge[Direct](X)(Y)
\Edge[Direct](Z)(X)
\Edge[Direct](Z)(Y)
\end{tikzpicture}

\caption{DAG showing common parent confounding where \textit{Z} is the parent of both \textit{X} and \textit{y}.  }
\label{fig1}
\end{figure}
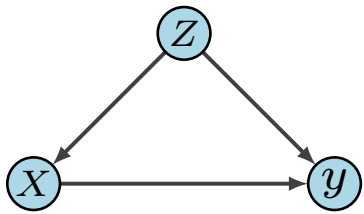

In causal inference, we are interested in estimating if (and to what extent) an event (or feature) causes an outcome of interest.  For example, do extra tutorials cause an increase in academic performance?  Consider Figure \ref{fig1} below which is a directed acyclic graph (DAG). This represents the true data generating process, the true causal structure in the data.  Let \textit{y} be the outcome of interest, say student performance, and let \textit{X} be extra tutorials.  Say we are interested in estimating the causal effect of extra tutorials (\textit{X}) on student performance (\textit{y}).  However, there is a third variable,  \textit{Z}, which in this case acts as a confounder, a parent, of both \textit{X} and \textit{y}.  Let's say \textit{Z} refers to personal motivation of the student; the inner drive to succeed.  What this specific causal graph is telling us is that the variable \textit{Z} is a cause of both attending the extra-tutorials and the performance:  more motivated and driven students will both perform better (\textit{y}) \textit{and} use every opportunity to get better, by attending extra-tutorials (\textit{X)}.  Therefore, if we are trying to estimate the true causal effect of extra-tutorials (\textit{X}) on performance \textit{y}, if we merely estimate the causal effect of \textit{X} on \textit{y}, there will be bias because of the confounding variable \textit{Z}. 

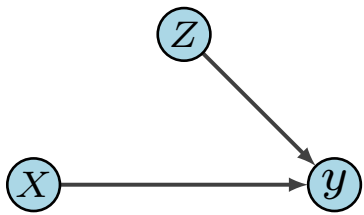
\begin{figure}[h!]
\centering
\renewcommand{\figurename}{\textbf{Fig.}}
\begin{tikzpicture}
\Vertex[x=1,size = 0.7,label = $X$, fontscale = 2]{X}
\Vertex[size =0.7, x=5,label = $y$, fontscale =2.5]{Y}
\Vertex[size=0.7,x=3,y=2,fontscale = 2,label = $Z$]{Z}
\Edge[Direct](X)(Y)

\Edge[Direct](Z)(Y)
\end{tikzpicture}

\caption{DAG showing adjusting for \textit{Z}.  }
\label{fig21}
\end{figure}

To estimate only the causal effect of \textit{X} on \textit{y}, we need to remove the effect of \textit{Z} on \textit{X}.  We can do this via Pearl's Do-Calculus which is to block the causal path from \textit{X} to \textit{y}.  We perform do(\textit{X}) which is to manipulate \textit{X} to see its effect on \textit{y}.  This can be performed practically by running randomized experiments or carrying out matching or regression techniques. 
 This is to essentially adjustment or control for, the variable \textit{Z}, i.e. fix it so that it no longer has an effect on \textit{X}.  This essentially breaks any relationship that \textit{X} has with \textit{Z}, as shown in Figure \ref{fig21}.  The detailed description of causal inference and do-calculus is beyond the scope of this study, but this example is presented to show how causal inference can be used to estimate true causal effects.

The point of the the above discussion is to show the the power of having a \textit{causal graph} and to know the true causal structure and true data generating process.  If we didn't have such a graph, we might not have known that \textit{Z} is a confounding feature that needs to be adjusted for using causal inference methods.  Knowing the causal structure in the data shines light on exactly how to apply causal inference methods to estimate true causal effects.

Three basic causal structures exist.  One is the fork which is shown in the example we just discussed, where we have a confounder or parent of two other variables:  \textit{X} $\leftarrow$ \textit{Z} $\rightarrow$ \textit{y}.  Another is a chain structure where one feature causes a second that causes the third:  \textit{X} $\rightarrow$ \textit{Z} $\rightarrow$ \textit{y}. The last is a collider where two independent variables cause a third: \textit{X} $\rightarrow$ \textit{Z} $\leftarrow$ \textit{y}.  These are three distinctly different causal structures.  However, if you were to obtain a dataset with these three variables you would not know how they are causally related.  If we do not know the causal structure, then estimating the treatment effects of one feature on another would most likely be biased.

\subsection{Causal Discovery}
Therefore if we desire to more accurately measure treatment (causal) effects by applying causal inference methods, we must know what the data generating process (causal graph) looks like.  This is where causal discovery comes in.  Causal discovery aims to take a real-life dataset where we do not know the complete causal structure (or graph), and generate the causal graph.  This causal graph can then guide us into a host of causal inference methods that could add much value to the problem we are trying to solve.  

Causal discovery (CD) was first proposed in 2000 to study gene expression \cite{spirtes2000b}.  Over the last twenty years there has been widespread use of these methods as well as a large increase in the number of CD algorithms developed \cite{Glymour2019}.    CD methods generally fall into four categories: constraint-based methods, score-based methods, functional causal discovery, and gradient-based methods \cite{Glymour2019,Molak2023}. In this study we limit our methods to the first two.  The aim of this study is to present the potential benefits of and downsides to, using causal discovery for student performance.  In future studies we will compare the different methods in more detail.  It is vital to include here that causal discovery works optimally when a domain expert or previous domain knowledge is available to model the causal graph.   We next consider two common CD methods that are mainly used in this study.

\subsubsection{Constraint-based methods}
This method is based on the concept of conditional independance tests.  For example, consider two variables x and y.  If we determine that x and y are correlated using statistical tests, then it could either mean that x causes y (x $\rightarrow$ y), or y causes x (y $\rightarrow$ x), or a third confounding variable exists such that x $\leftarrow$ z $\rightarrow$ x.  Therefore constraint-based methods use conditional independence tests to estimate potential causal relations.  However, the limitation is that at best, these methods produce a Markov equivalence class (MEC): where multiple causal graphs produce the same conditional independence.  In other words, these tests can at best discover a class of causal graphs that produce the same condition independence.  However, the positive side is that we can then look at the potential graphs and use logic and domain knowledge to identify the true graph. More detail on this can be found in \cite{Glymour2019,Molak2023}.  The PC algorithm is an example of a constraint-based method and perhaps traditionally the most commonly used CD method.

\subsubsection{Score-based methods}
Whereas the PC method (constraint-based) starts with a fully connected graph and uses conditional independence to remove edges, score-based methods such as greedy-equivalence-search method (GES), start with a blank graph and add and remove causal edges, again based on statistical tests. Each generated graph is an attempt to fit the observational data in the best way.  To each graph GES assigns a score and the best score aims to represent the most likely causal relationship in the data.  

\section{Counterfactual Analysis}
\label{PCM}
 A counterfactual represents a hypothetical scenario where one or more factors are changed while keeping everything else the same.  The aim is to determine how an output will change in a counterfactual scenario.   This idea is key to providing personalized recommendations to students or to any other field of interest. 
 
 The aim of counterfactual analysis is to predict how an outcome would have changed had we manipulated the input causal feature: what would we need to do to the causal inputs in order to obtain a desirable output.  In the context of student performance, how would an individual student need to change some causal input in order to switch from being at-risk to passing?  

In this study we follow the counterfactual method developed by Judea Pearl \cite{Pearl2017}, and not those presented as counterfactual explanations in the explainable machine learning literature \cite{Molnar2019}.  Reasons for this are beyond the scope of this study but can be found in \cite{SmithArxiv2023}.  Note that counterfactual analysis is performed after we have performed causal discovery to find the causal structure. 

Pearl's counterfactual method follows the following steps:
\begin{enumerate}
    \item Abduction:  Once we have the true causal structure and have estimated the structural causal models, we use existing measured data for an individual case of interest, and compute all exogenous (noise) inputs.  Computing the noise variables for individual cases is a vital aspect of counterfactual analysis.
    \item Action:  After computing the noise variables, we intervene (perform do-calculus) on the causal input variable of interest.  This we sever the relationship that the input causal variable had with its' parents \cite{Pearl2017,SmithArxiv2023}.  Intervening on a variable means all inputs into that variable no longer affect it.

    \item Prediction:  We now choose counterfactual inputs and using the noise variables calculated for the individual case, we predict counterfactual outcomes.   Ultimately, we want to identify a desirable outcome (such as a student passing the course) and the necessary counterfactual inputs.

\end{enumerate}

For a detailed description of Pearl's counterfactual analysis see \cite{Pearl2017,SmithArxiv2023}. 

\subsection{Counterfactual Requirements: Causal and Actionable}
\label{actionable}
In order for counterfactual recommendations to be meaningful, they need to be causal and actionable.  Causality is satisfied by the causal discovery step where we uncover the causal structure.   However, even though features may cause another feature, they might not be actionable.  For example, we may find that age is an important cause for an output.  This type of feature is not actionable: we can't physically manipulate someone's age to change the output.  Or we may find a causal feature has already occurred and you cannot go back in time to manipulate that feature.

What can we do in these cases?  We can find proxy factors that are related.  For example, we can ask ourselves, why does the age variable cause this output?  Is it experience in something else that is the true cause?  Or the feature that has already occurred, is there something that it is currently related to that might also be causal?

\subsection{Causal Inference Steps}
Based on the above, we identify the causal inference steps needed to provide personalized counterfactual recommendations for students (or for similar cases).

\begin{enumerate}
    \item Perform causal discovery: 
 uncover/discover the true data generating process, the causal structure in the data, the DAG.
    \item Identify the causal question you want to answer. Average treatment effect and/or counterfactual?
    \item Perform do-calculus (adjust, control, etc.) to estimate average treatment effects.
    \item Perform counterfactual analysis to provide personalized recommendations.
\end{enumerate}

\section{Methods}
The above was applied in two parts: first to synthetic data (Section \ref{synth_stud}) and then to real-life data (Section \ref{realLife}).   A synthetic causal structure (and data) was generated  with the aim of seeing how well the causal discovery methods perform on data with a \textbf{ground truth} causal structure. How well do these methods discover the true causal structure?  We study the PC and GES algorithms here.

Second, CD and counterfactual analysis were then applied to real-life data. The major challenge with real-life data is that we have no ground truth causal structure.  We use the PC algorithm for CD here.

\subsection{Synthetic Student Causal Structure}
\label{synth_stud}
Synthetic data was generated mimicking student data as shown in Table \ref{dataset1}.  The simulated data represents student information such as grades, gender, age etc.

\begin{table}[h!]
\centering
\caption{Description of features used to generate the synthetic student data for the chain causal structure.}
\vspace{0.25cm}
\begin{tabular}{|l|l|l|}
\hline

Feature & Description & Statistics \\
\hline
x1      & Grades      & $\mu$=50, $\sigma$=5       \\
x2      & Age         & $\mu$=20, $\sigma$=1       \\
x3      & Grades      & $\mu$=45, $\sigma$=6       \\
x4      & Gender      & \textit{p} = 0.6        \\
x5      & Bursary     &\textit{p} = 0.3        \\
x6      & Grades      & $\mu$=70, $\sigma$=5       \\
x7      & Grades      & $\mu$=50, $\sigma$=5     \\ \hline
\end{tabular}
\label{dataset1}
\end{table}

\begin{figure}[h!]
\renewcommand{\figurename}{\textbf{Fig.}}
\centering
\begin{tikzpicture}
\Vertex[x=0.8,size = 0.7,y=2,label = $e_1$, fontscale = 2]{e1}
\Vertex[x=1,size = 0.7,label = $x_7$, fontscale = 2]{x7}
\Vertex[size =0.7, x=5,label = $y$, fontscale =2.5]{Y}
\Vertex[size =0.7, x=5,y=1.5,label = $X$, fontscale =2.5]{X}
\Vertex[size=0.7,x=3,y=2,fontscale = 2,label = $x_3$]{x3}
\Vertex[size=0.7,x=7,y = 1.5,label = $e_2$,fontscale = 2]{e2}
\Edge[Direct](X)(Y)
\Edge[Direct](e2)(Y)
\Edge[Direct](e1)(x3)
\Edge[Direct](x7)(x3)
\Edge[Direct](x3)(Y)
\end{tikzpicture}
\caption{Causal structure (data generating process) for synthetic student data, based on the chain causal structure.}
\label{synth1}
\end{figure}
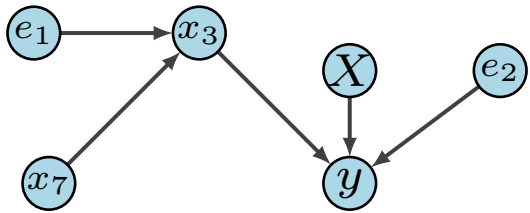

The \textit{causal structure} was then then generated according the chain causal structure shown in Figure \ref{synth1} ($x_7$ $\rightarrow{x_3}$ $\rightarrow{y}$) and the structural causal model (SCM) shown in Equations \ref{exp100a} and \ref{exp100b}. This data therefore simulates a final grade \textit{y} as a function of the other variables.

\vspace{2cm}

\begin{align}  
&x_3= \beta_0 x_7 + e_1  \label{exp100a} \\
&y = \beta_0 + \beta_1 x_1 + \beta_2 x_2 + \beta_3 x_3 + \beta_4 x_4 + \beta_5 x_5 + \beta_6 x_6 + e_2  \label{exp100b}
\end{align}

We also include two errors in the data generating process, $e_1$ = $\sim \mathcal{N}(0, 1)$ and $e_2$ = $\sim \mathcal{N}(0, 2)$.  The $\beta$ coefficients from 0 to 6 were arbitrarily selected as follows: 0.4, 0.6, 0.4, 0.6, 0.7, 0.4, and 0.4. As can be seen from Figure \ref{synth1}, it is important in the data generating process, to make sure the errors are different, otherwise we introduce confounding.  That is, if the error into $x_3$ and into \textit{y} are from the same distribution, it acts as a parent of both $x_3$ and \textit{y}, introducing problems when applying causal discovery.  Errors (or noise) in statistical models refer to all other inputs into the data that are not accounted for in the measured features.

\subsection{Real-life Data}
\label{realLife}
The real-life data was taken from Smith et al. \cite{BevanSmith2022a} which records student performance data for a first year engineering mechanics course.  All details can be found there.  The dataset comprised 39 input features such as gender, age, ethnicity, degree of study, high school grades, first year engineering grades etc. and one target feature, the final engineering mechanics grade.  There were 878 observations, i.e. students, from which we captured the data thereby utilizing a dataset of with shape (878,40).

\subsection{Causal Discovery Methods}
As mentioned earlier, the aim was to study how using CD can be used for personalized student recommendations.  We mainly applied the PC algorithm.  Future studies will include other methods.

\section{Results}
\subsection{CD on Synthetic Student Data}
\begin{figure}
    \centering
    \includegraphics[scale=0.3]{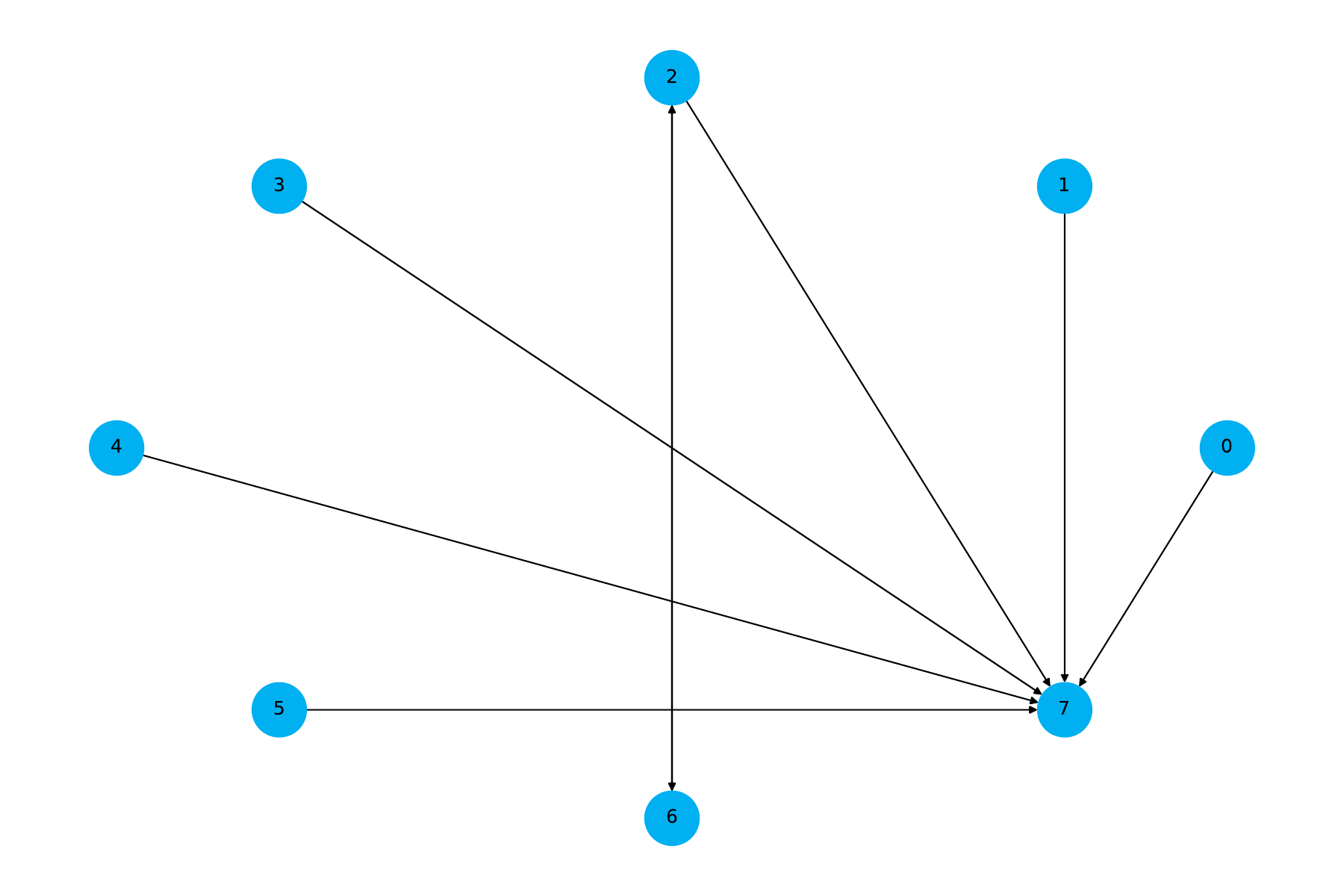} 
    \caption{Causal graph of synthetic student data using the PC and GES algorithms (identical).}
    \label{fig:pc_synth1}
    
\end{figure}

The causal graph produced by the PC and GES algorithms were identical, as shown in the Figure \ref{fig:pc_synth1}.  The results closely match the ground truth causal graph in Figure \ref{synth1},  except for the relationship between x7 and x3, shown as 6 and 2 in the graph, respectively.  This can be attributed to the contraint-based methods being conditional probability methods that can at best produce Markov Equivalence Classes. 

\begin{figure}
    \centering
    \includegraphics[scale=0.3]{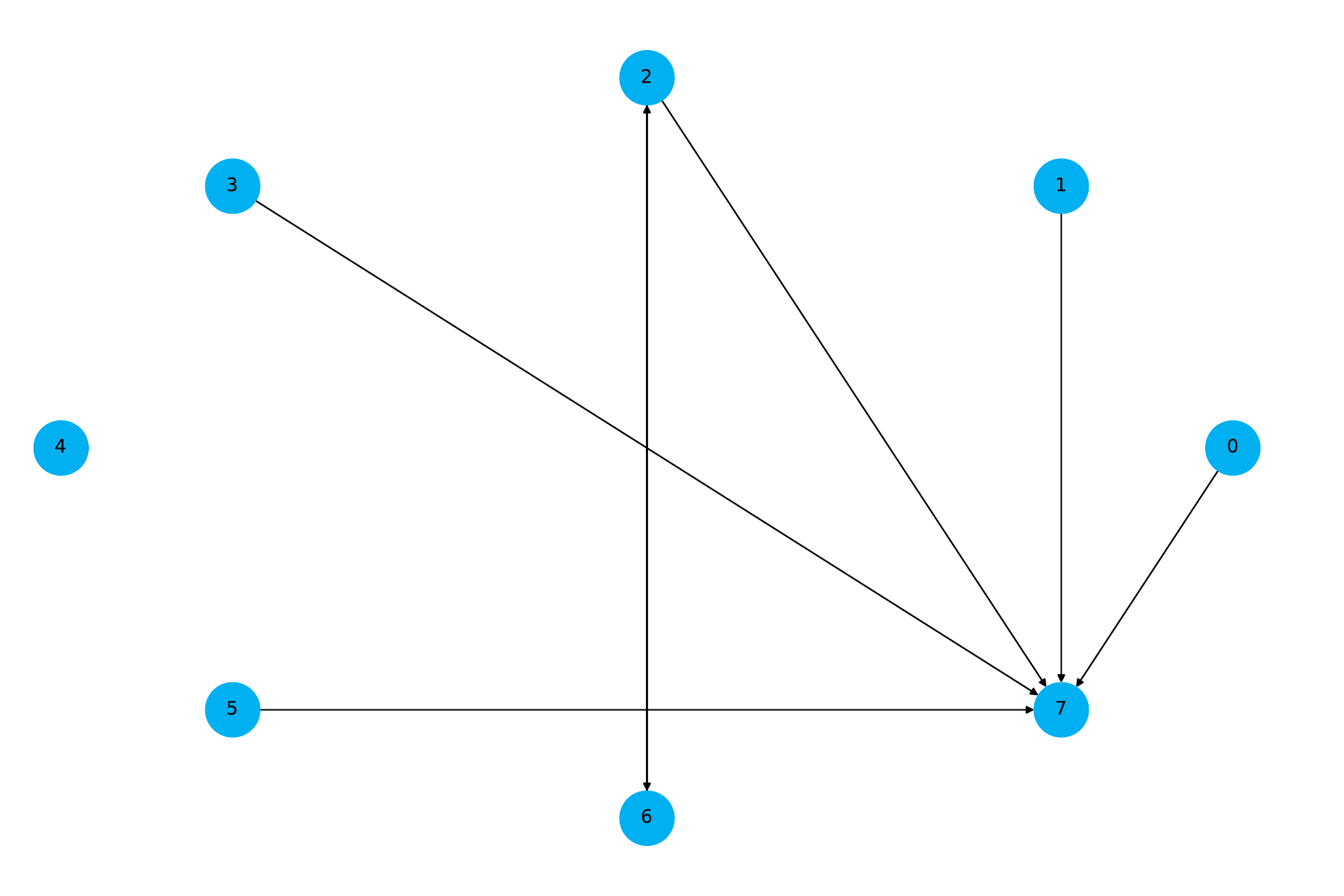} 
    \caption{Second causal graph of synthetic student data using the PC and GES algorithms (identical).}
    \label{fig:ges_synth2}
    
\end{figure}

To show how these methods are somewhat volatile, PC and GES were again applied to the dataset (without seed) to produce the DAG in Figure \ref{fig:ges_synth2}.  Here we can see that the algorithms (again producing identical DAGs) estimate no relationship between x5 (node 4) and y (node 7).  This shows that even for relatively simple causal structures, the models can generate incorrect causal structures.  However, it was further found that as we increased the sample size to much larger samples, the models produced correct DAGs each time.

Therefore what we can see from the synthetic data is that as the data size grows the models become more trustworthy but at low sample sizes, incorrect DAGs tend to be produced.

\subsection{Student Performance Data}

\begin{figure}
    \centering
    \includegraphics[scale=0.3]{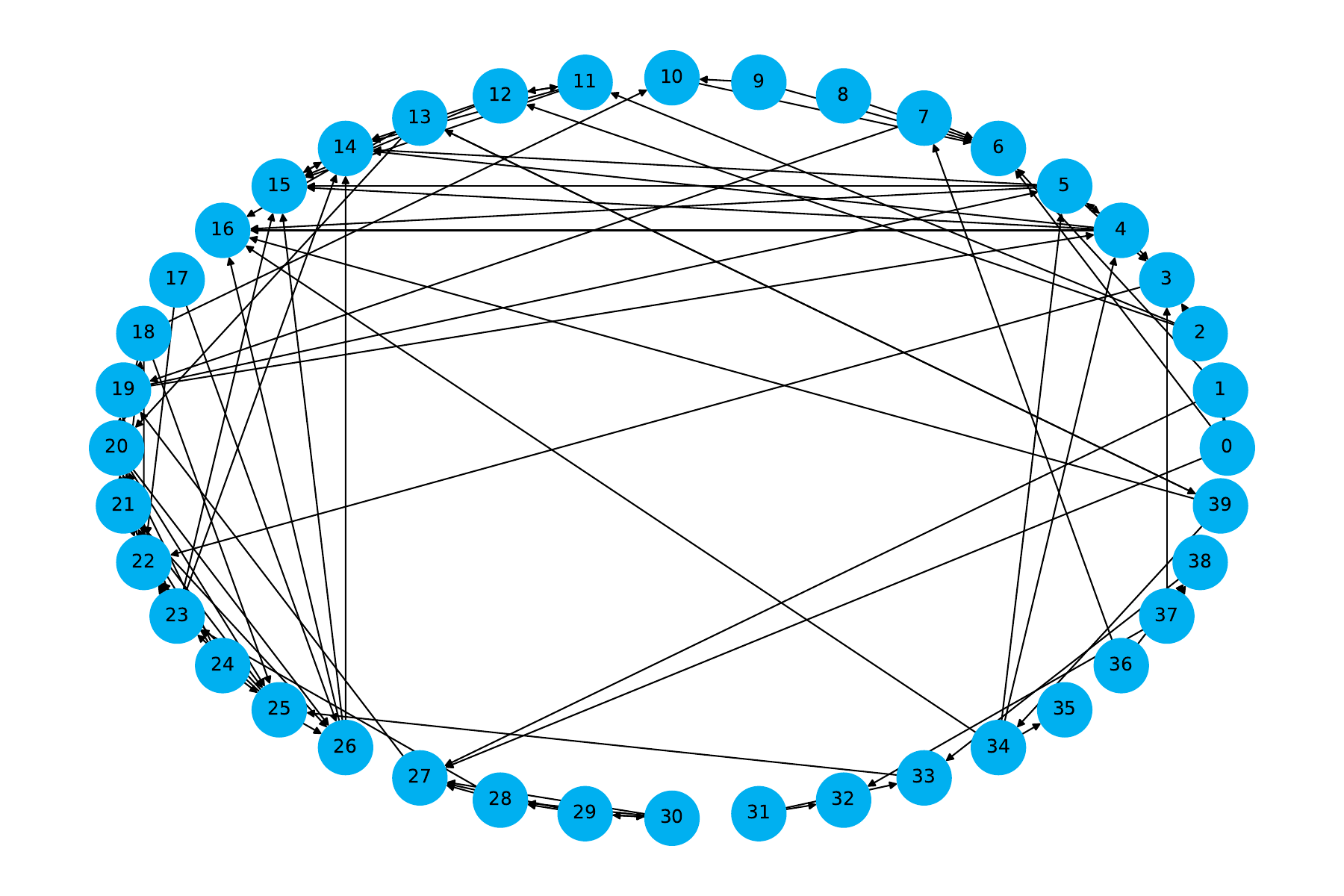} 
    \caption{Causal graph of student performance data generated using PC algorithm}
    \label{fig:pc_student}
    
\end{figure}

Here we apply CD and counterfactual analysis to real-life student data. Using the PC algorithm, we obtained the causal structure shown in Figure \ref{fig:pc_student}.  The dataset comprised 39 input features and 1 target feature.  The DAG represents all 40 features.

\begin{figure}
    \centering
    \includegraphics[scale=0.3]{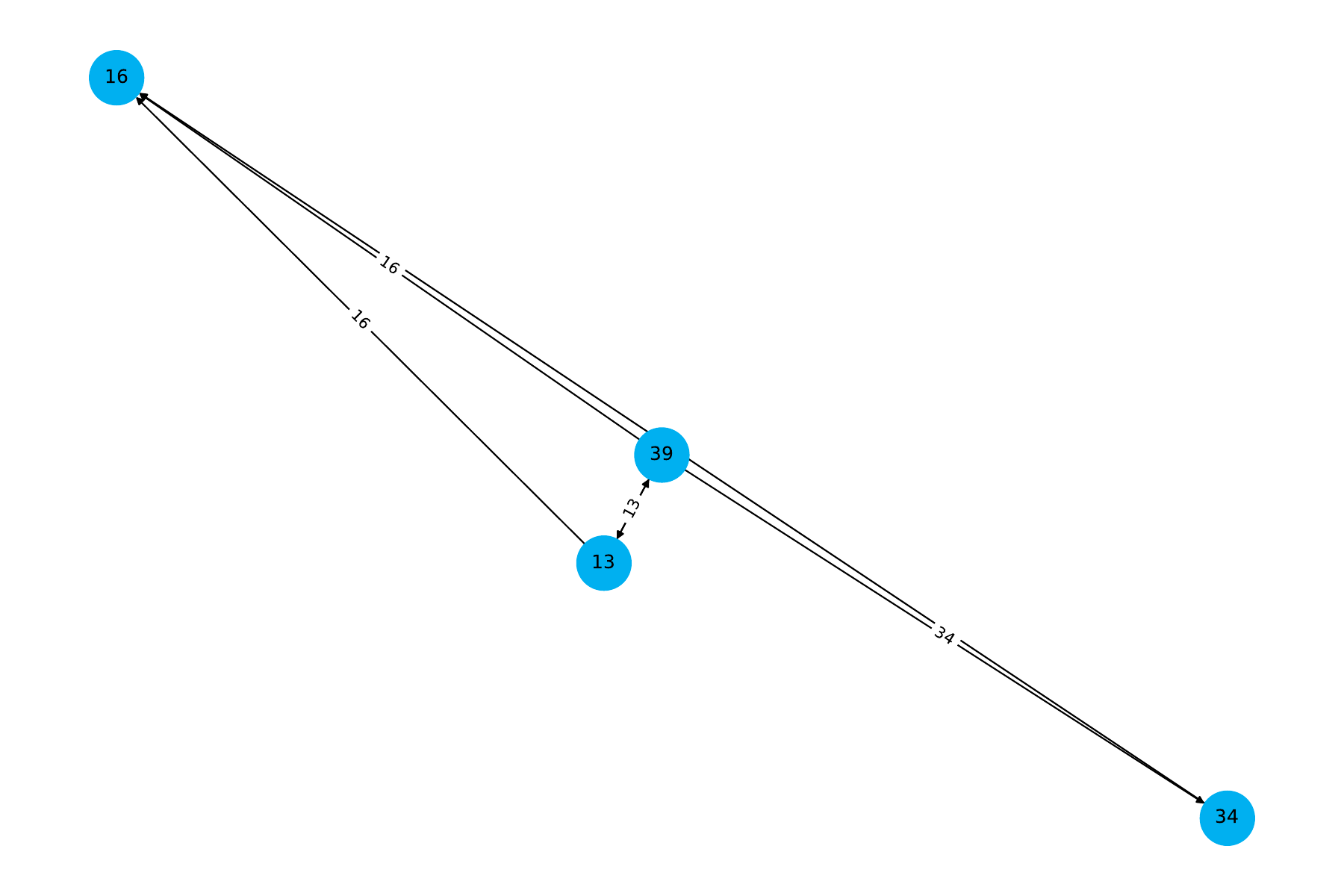} 
    \caption{Causal graph isolating the feature of interest. The graph was generated without incorporating priori constraints.  Impossible causal directions are included here. }
    \label{fig:pc_student391}
    
\end{figure}

It is more valuable to isolate the feature we are interested in, namely the final student performance, node 39, to see what is directly influencing that feature. Note that Python indexes data from 0 and not 1, hence the target variable being node 39 and not 40. The reduced DAG is shown in Figure \ref{fig:pc_student391}.  From this initial DAG produced by the PC algorithm, we can see that a few relationships don't make sense.  First of all, the final student performance, node 39, has arrows pointing \textbf{to} other features, nodes 16 and 34.  This is impossible since those features occurred in time before node 39. Nodes 16 and 34 refer to grades that occurred either earlier in the semester or previously in high school.  In actual fact, no features in the dataset occurred \textit{after} the target feature.  This is helpful because it allows us to introduce a constraint into the PC algorithm called priori knowledge.  Using this constraint we are able to educate the algorithm as to which causal directions are allowed and which are forbidden as it generates the DAG.

\subsubsection{Priori Knowledge}
\begin{figure}
    \centering
    \includegraphics[scale=0.3]{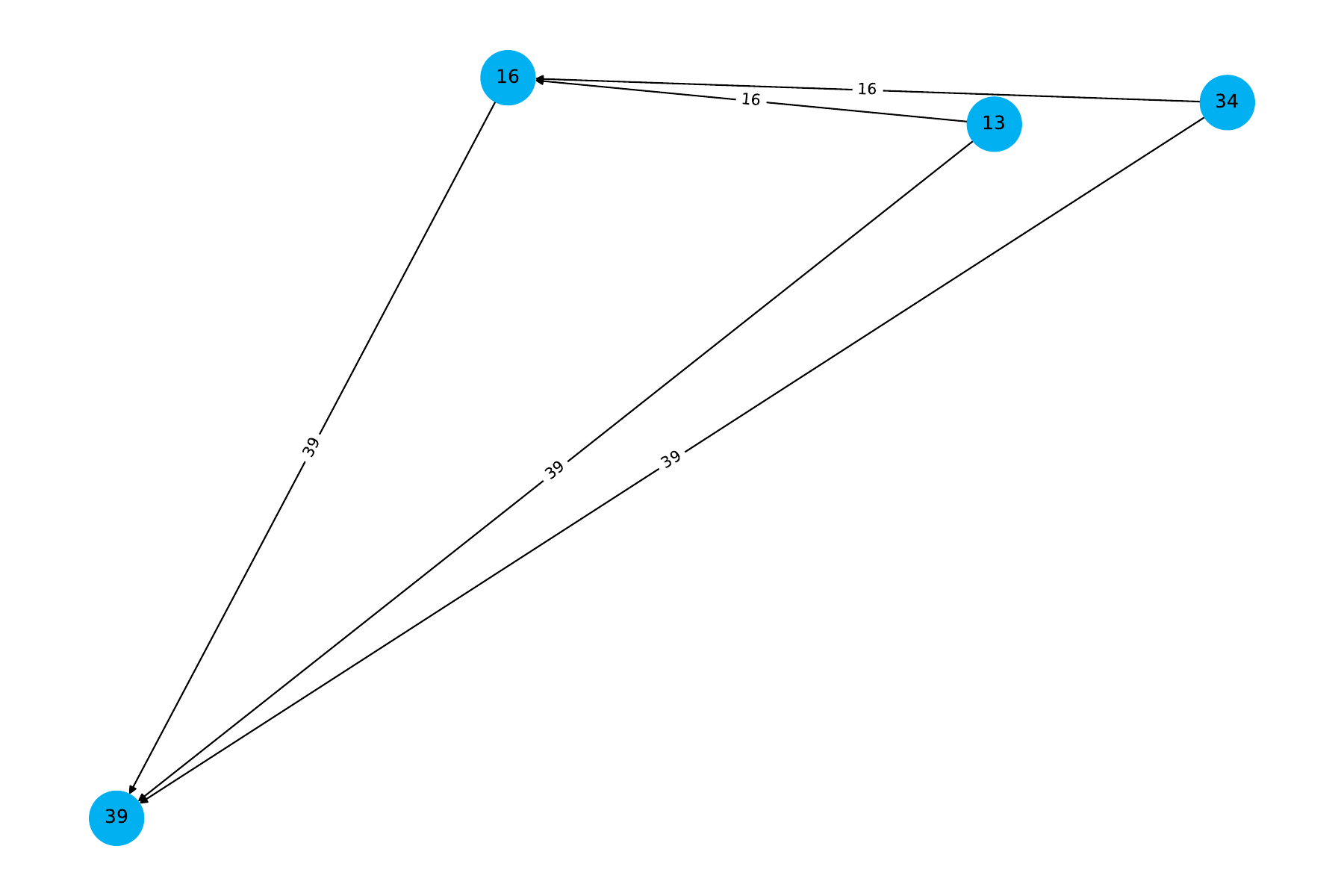} 
    \caption{Causal graph of only feature of interest data generated using PC algorithm}
    \label{fig:pc_student3}
    
\end{figure}

Because node 39 occurs temporally after all the features in the dataset, arrows \textit{from node 39 to} any other feature was forbidden and the PC algorithm was applied again and the DAG in Figure \ref{fig:pc_student3} was generated.  Here we see all the arrows pointing to node 39 or to each other, but no arrows point away from node 39, which is correct. 
 This also shows how a domain expert might introduce causal relationships into the model.

Analyzing the actual features from Figure \ref{fig:pc_student3}, we see three causal features into the target feature (node 39). 
 Nodes 13 and 16 refers to the mechanics and math test grades occurring earlier in the semester, and node 34 refers to mathematics grades in high school.  These results are plausible.  It makes perfect sense that the previous mechanics test grade (node 13) is a cause of the final mechanics grade.  It further makes sense that mathematics has a causal influence on engineering mechanics since it contains a substantial math component, namely linear algebra and calculus.  What the DAG further suggests is that the mechanics grades (13) and high school math act as confounders, parents, of the math test (node 16) and final mechanics grades.  This suggests that mathematical ability is a cause of both the previous mechanics grades and the final grades.  This is plausible.     

As an initial observation, this suggests that to be successful at engineering mechanics requires high performance in both high school and first year engineering mathematics.  Therefore if general, if we are advising a student to improve his/her mechanics marks, then we can advise them improve their earlier mechanics and mathematics grades.  That is very general.  However, the causal structure is not an end in itself.  We now have insight into how to perform causal inference to estimate the true causal (treatment) effects of another feature on the final mechanics grades (Section \ref{treatEffects}) and to perform counterfactual analysis (\ref{countRecomm}).

\subsection{Estimating Treatment Effects}
\label{treatEffects}
We could now ask causal questions about any of the causal inputs into node 39.  Let's consider how node 16, the math test, is causally influencing the final mechanics grade (node 39).  To do this we need to perform causal inference: control/adjust for the confounders, nodes 13 and 34.  Once we control for those features, we can then estimate the true treatment effect of the previous math test on the final grade.  Discovering the DAG has helped us identify which features to control for to estimate the true treatment effect of any of those variables on the final grades.  This is shown next. 

This can be done using a regression analysis which would include nodes 16, 13 and 34 as the independent variables and node 39 as the dependent.  Note that if we only include node 16 as the independent variable, then the regression model will estimate a biased coefficient (treatment effect).  Figure \ref{fig:reg1} below shows that all three input variables are significant (\textit{p} < 0.05), but more importantly, the coefficient of 0.199 for March.Test.MATH1014 (math test, node 16) refers to the true treatment effect of math test on final mechanics grades. 
Had we \textit{not} included the other two variables, the estimated treatment effect of the math test would have been approximately 0.6 which would have overestimated the causal effect.  We now have the true causal effect of math test on final mechanics grades.

\begin{figure}
    \centering
    \includegraphics[scale=0.7]{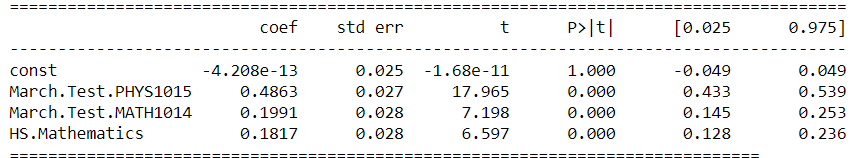} 
    \caption{Regression results for estimating causal effect of mechanics test on final mechanics grades.}
    \label{fig:reg1}
    
\end{figure}

The coefficient of 0.199 indicates that for every increase of one unit of the math test, there is approximately a 0.2 increase in the final mechanics grades.  A very important point to note is that these are average treatment effects.  We could therefore advise an individual student that the better you perform in the math test, the better you will perform in the final grade; but that would not necessarily be a personal recommendation.  For this, we need to turn to counterfactual analysis, shown next.  Although not performed here, we could easily use causal inference methods to also estimate the true causal effects of mechanics test (node 13) or high school math (node 34) on final grades. For example, we can see that node 13 has causal flow directly to node 39 as well as through node 16.  This includes a chain type causal structure which requires specific causal inference methods.

\subsection{Counterfactual Recommendation}
\label{countRecomm}
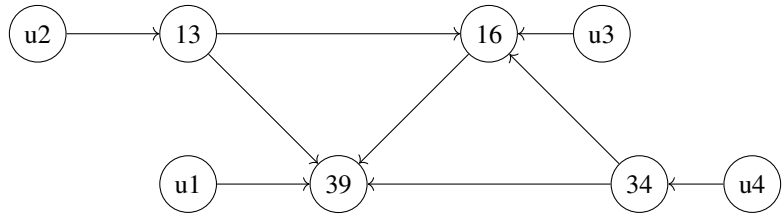
\begin{figure}
\centering
\begin{tikzpicture}
    \node[circle, draw] (39) at (0, 0) {39};
    \node[circle, draw] (16) at (2, 2) {16};
    \node[circle, draw] (13) at (-2, 2) {13};
    \node[circle, draw] (34) at (4, 0) {34};
    \node[circle, draw] (u1) at (-2,0) {u1};
    \node[circle, draw] (u2) at (-4,2) {u2};
    \node[circle, draw] (u3) at (3.5,2) {u3};
    \node[circle, draw] (u4) at (5.5,0) {u4};
    
    \draw[->] (16) -- (39);
    \draw[->] (13) -- (16);
    \draw[->] (13) -- (39);
    \draw[->] (34) -- (16);
    \draw[->] (34) -- (39);
    \draw[->] (u1) -- (39);
    \draw[->] (u2) -- (13);
    \draw[->] (u3) -- (16);
    \draw[->] (u4) -- (34);
\end{tikzpicture}
\caption{Causal graph reproduced for counterfactual analysis. Noise terms are included here.}
\label{cf1}
\end{figure}

The previous section was valuable for estimating average treatment effect.  The aim though, is to not only provide average treatment effect information, but to provide personal recommendations.  Here we turn to Pearl's counterfactual method discussed in Section \ref{PCM}.  To illustrate how we generate a counterfactual, consider Figure \ref{cf1} which is a reproduction of Figure \ref{fig:pc_student3}, but now with important noise terms included.  Noise terms refer to all other causes of variance in a variable that are not accounted for in the measured/observed variables.  They are vital for generating counterfactuals, because (1) they should be unique to an individual student and (2)  the noise should be fixed while changing the causal features.  The assumption is that the noise remains invariant while the counterfactuals are changed.  

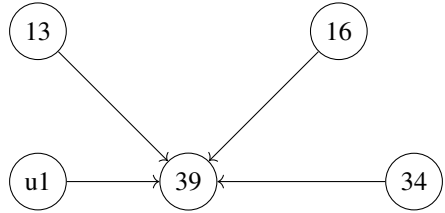
\begin{figure}
\centering
\begin{tikzpicture}
    \node[circle, draw] (39) at (0, 0) {39};
    \node[circle, draw] (16) at (2, 2) {16};
    \node[circle, draw] (13) at (-2, 2) {13};
    \node[circle, draw] (34) at (3, 0) {34};
    \node[circle, draw] (u1) at (-2,0) {u1};

    \draw[->] (16) -- (39);
   
    \draw[->] (13) -- (39);
    
    \draw[->] (34) -- (39);
    \draw[->] (u1) -- (39);

\end{tikzpicture}
\caption{Causal graph when manipulating nodes 13, 16 and 34.  Parent relationships are severed.}
\label{cf2}
\end{figure}

Step 1 of Pearl's method is to perform Abduction. Because we aim to manipulate all three nodes (13, 16 and 34), that is, generate counterfactuals using all three nodes, we sever their relationships with their parents.  Therefore, in that case, we don't need to calculate their noise terms, only the noise term $u_1$, feeding into node 39 (the final mechanics grade).   
After severing the parent relationship, we end up with the causal graph in Figure \ref{cf2} and the structural causal model (SCM) in Equation \ref{exp200b}.

\begin{equation}
y =  c_1 n_{13} + c_2 n_{34} + c_3 n_{16} + u_1    \\
 \label{exp200b}
\end{equation}

In order to calculate the noise term $u_1$ for a particular student, we need to input the observed data into the SCM.  We know \textit{y}, $n_{13}$, $n_{34}$, $n_{16}$ as well as the coefficients (see Figure \ref{fig:reg1}).  We input that into the equation and calculate $u_1$.  This is the first step of Abduction.

To illustrate this, we select actual observed data for a random student from the population who is predicted to fail the course, i.e. at-risk.  Note that all data has been normalized for computation and anonymizing purposes.  See the student's normalized data in Table \ref{atriskstudent}.

\begin{table}[ht]
\centering
\begin{tabular}{|c|c|c|c|}
\hline
Mechanics Test (13) & Math Test (16) & High School Math (34) & Final Grade (39) \\
\hline
0.06& -2.57 & -0.365 & -1.29 \\
\hline
\end{tabular}

\caption{Randomly selected at-risk student data.}
\label{atriskstudent}
\end{table}

We now feed the observed data from Table \ref{atriskstudent} into Equation 3 and compute $u_1$ to be -0.763.

\begin{align}
    y &= c_1 n_{13} + c_2 n_{34} + c_3 n_{16} + \bm{u_1} \\
-1.29 &= 0.19 \cdot -2.57 + 0.486 \cdot 0.06 + 0.187 \cdot (-0.365) + \bm{u_1}\\
\bm{u_1} &= -0.763
\end{align}

We have now computed the noise term for this specific student.  We can now apply the final step of Prediction: to calculate counterfactual quantities.  We use the noise term and input counterfactual input features to compute a counterfactual output \textit{y}.  Recall that the aim is to manipulate the input causal features to change the student's results from failing to passing.   This is an optimization exercise which goes beyond the scope of this study.  The point is that we have three options (nodes 13, 16, 34) and multiple combinations of those input features, to change in order to affect the output.  Due to the data being normalized, for our dataset, the passing value in normalized terms was z-score = -0.901 (for 50\%).  The student we selected has a z-score = -1.29 which is below passing.  Therefore in order to increase this student's mechanics grade from -1.29 to -0.901, we could, for example, keep everything the same and only change the math test (16) from a z-score of 0.06 to  approximately 0.9.  As you can see there are multiple options and combinations that could achieve the desired output of improving the student's grade from -1.29 to -0.901.  This is an optimization exercise.   We can therefore use these results to advise the student on how to improve in order to pass the course.

\section{Discussion}
The study presented in this paper explored the application of causal discovery methods and counterfactual analysis to student performance data. Two main contributions were made: (1) the use of causal discovery to reveal the underlying causal structure of the data, enabling informed causal inference, and (2) the implementation of Pearl's counterfactual method for generating personalized recommendations.

There are however, some limitations of the current study.  Although the features that were selected using causal discovery as causal inputs (nodes 13, 16 and 34) are indeed plausible, they are not actionable.  This means that the student cannot actually act on those specific features.  These three features have already occurred and so the student cannot go back and change his/her mechanics and math test grades or high school math grades.  Nevertheless, as mentioned in Section \ref{actionable}, we can use proxies for the actual feature and advise the student in this case to focus a substantial amount of time on their math skills, specifically those related to mechanics. Identifying actionable features or proxy features is important. We could in future studies, make sure we included later assessments that could be actionable.

Further challenges to using causal discovery is that we do not know the ground truth causal structure and SCMs.  For this we need domain experts who might know the true causal relationships.  Recall that at best, the PC and GES algorithms produce Markov Equivalence Classes.  A domain expert can then provide known causal relationships for the CD algorithm.  Unless there are obvious causal relationships that are easy to spot, it is advised that CD algorithms not be used on their own and applied without human experts guiding them. 

Another major challenge and a potential huge area for future work is that this study focuses on historical data.  It has not been applied in real-time.  That is, we have not applied this in real-time yet to a student early in the semester to provide personal recommendations to him/her on how to switch from being at-risk to passing.  This is perhaps the greatest challenge in counterfactual recommendations.  Say we go through the steps outlined in this study and generate counterfactual recommendations for a student.  For example, say a month into the semester we identify an at-risk student: meaning we \textit{predict} that this student will fail at the end of the semester.  We then use all relevant data and generate counterfactual recommendations that if followed, \textit{predict} that the student will pass.  The challenge here is two-fold.  The first is, how do we ensure that he/she follows the recommendation fully? The student may follow it for some time and then fade away. 
 Then it is difficult to know if our recommendations worked.   The second challenge is, how do we keep all other things constant?  A counterfactual assumes that we only change that thing (or those things), while all else remains the same.  Say we recommended that the student needs to improve her mathematics skills.  Then she begins focusing on that but perhaps by focusing on that, she focuses less on other important things.  This might now change the data distribution (i.e the causal structure) that we assumed when performing causal discovery and counterfactual analysis.  This is therefore a challenging problem.  But we believe this is a very promising set of methods that can substantially help students.  We need to invest more focus and study on how to solve these problems.  Nevertheless, we believe that even though it is challenging to validate how effective these methods are, being able to identify the root cause of a student's poor performance and knowing that it is causal, is still of great value.  This field is highly promising, not only for students but for any field where we desire to know the causal influences and generate counterfactual recommendations.

\bibliographystyle{splncs04}
\bibliography{ReferencesCurrent2}
\end{document}